\documentstyle[psfig,astrobib]{mn}
 
\def\be{\begin{equation}}
\def\ee{\end{equation}}
\def\ni{\noindent}
\def\bfig{\begin{figure}}
\def\efig{\end{figure}}
\def\bfigw{\begin{figure*}}
\def\efigw{\end{figure*}}
\def\etal{et al.~}
\def\etalc{et al.}

\def\dd{\hbox{\rm d}}
\def\spose#1{\hbox to 0pt{#1\hss}}
\def\approxlt{\mathrel{\spose{\lower 3pt\hbox{$\sim$}}
	\raise 2.0pt\hbox{$<$}}}
\def\approxgt{\mathrel{\spose{\lower 3pt\hbox{$\sim$}}
	\raise 2.0pt\hbox{$>$}}}
\def\aap{A\&A}

\def\aj{AJ}
\def\apj{ApJ}
\def\apjl{ApJL}

\def\araa{ARA\&A}
\def\mnras{MNRAS}
\def\nat{Nat}

\def\procspie{SPIE Proc.}
\def\ISO{{\em ISO }}
\def\ISOPHOT{{\em ISOPHOT }}
\def\IRAS{{\em IRAS }}
\def\cm{{\rm\thinspace cm}}
\def\mm{{\rm\thinspace mm}}
\def\erg{{\rm\thinspace erg}}
\def\Jy{{\rm\thinspace Jy}}
\def\mJy{{\rm\thinspace mJy}}

\def\Hz{{\rm\thinspace Hz}}
\def\g{{\rm\thinspace g}}
\def\K{{\rm\thinspace K}}

\def\keV{{\rm\thinspace keV}}

\def\m{{\rm\thinspace m}}
\def\Mpc{{\rm\thinspace Mpc}}
\def\s{{\rm\thinspace s}}

\def\sr{{\rm\thinspace sr}}

\def\W{{\rm\thinspace W}}
\def\col{\hbox{$\cm^{-2}$}}
\def\fluxerg{\hbox{$\erg\cm^{-2}\s^{-1}\,$}}
\def\fluxw{\hbox{$\W\m^{-2}\sr^{-1}\,$}}

\def\ergps{\mbox{$\erg\s^{-1}$}}

\def\pcm{\hbox{$\cm^{-3}\,$}}
\def\pcm3{\hbox{$\cm^{-3}\,$}}

\def\zmax{z_{{\it max}}}

\def\um{\hbox{$\mu {\rm m}$}}
\def\Ang{{\rm\thinspace \AA}}
\def\kband{$K$--\,band~}
\def\eg{e.g.}
\def\ie{i.e.}

\begin{document}
 
\title[Sub-mm/FIR Implications of Obscured AGN Models]
{Implications of an Obscured AGN Model for the X-ray Background 
at Sub-mm and Far Infra-Red Wavelengths} 
\author[K. F. Gunn and T. Shanks]
{Katherine F. Gunn\thanks{Present address: Department of Physics \&
Astronomy, University of Southampton, Highfield, Southampton, SO17 1BJ.}
 and Tom Shanks \\
Department of Physics, University of Durham, South Road, Durham, DH1 3LE.} 
 
\date{MNRAS - submitted.}
\pagerange{\pageref{firstpage}--\pageref{lastpage}}
\pubyear{1999}
 
\label{firstpage}

\maketitle
 
\begin{abstract}

Models invoking large populations of obscured AGN are known to provide
good fits to the spectrum of the X-ray background and the observed
soft and hard X-ray number counts.  An important consequence of these
models is that significant quantities of dust are required in order to
provide the obscuring medium, which will be heated by the nuclear
radiation from the AGN, and will radiate in the thermal infra-red.  We
therefore model the properties of the dust, and predict the
contribution of obscured AGN to the intensity of the far infra-red
background and the sub-mm source counts and redshift distribution, in
order to ascertain whether our models are constrained further by the
data available at these wavelengths.  Our conservative models predict
a contribution of between 5 and 15 per cent of the far infra-red
background intensity, whereas if we use more extreme values for
several parameters, this value may reach 33 per cent.  This suggests
that AGN may only form a significant minority of sub-mm sources, in
agreement with the results of spectroscopic follow-up.  Our models
thus show that there is no inconsistency between obscured AGN models
for the X-ray background and either sub-mm source counts or the
intensity of the far infra-red background.  We further propose that
obscured AGN may explain the sub-mm emission associated with Extremely
Red Objects (EROs).  Finally, we make predictions for the redshift
distribution of AGN sources detected in faint sub-mm surveys, which
should allow future tests of this obscured AGN model.

\end{abstract}
 
 
\section{Introduction}

Recently, enormous advances have been made in the field of sub-mm
astronomy, most notably with the commissioning of SCUBA, the Sub-mm
Common User Bolometer Array (\citeNP{Gear94}; \citeNP{Holland99}), on
the James Clark Maxwell Telescope (JCMT).  The $850\um$ surveys of
\nocite{SIB97} Smail, Ivison \& Blain (1997), \citeANP{Hughes98}
(1998) and \citeANP{Barger98} (1998) suggest that the sub-mm counts
show strong evidence of evolution.  The sources detected at this
wavelength appear to be dusty, high redshift galaxies.  However, the
origin of the emission is not clear, and these distant dusty galaxies
could be powered by either starburst or AGN activity.  The question of
which component dominates is highly important: if starbursts dominate,
then it would imply that a significant amount of star-formation has
been missed by studies which use ultra-violet techniques to claim that
the star-formation rate (SFR) decreases in galaxies at high redshift
\cite{Madau96}.  \citeANP{Metcalfe96} (1996) have previously suggested
that models which include dust with a monotonically {\it increasing}
SFR towards higher redshift can also give good fits to faint galaxy
counts and colours (see also \citeNP{Wang91}; \citeNP{Gronwall95};
\citeNP{Campos97}; \citeNP{Steidel99}).

It has also been known for some time that some ultraluminous infra-red
galaxies (ULIRGs), which are strongly star-forming, dusty galaxies,
also contain buried QSO nuclei.  Images of ULIRGs in polarized light
have shown highly anisotropic structure, such as that observed in the
\IRAS galaxy, F10214+4724, \cite{Lawrence93a}.  Similar
polarization structure is seen in Seyfert 2s and high redshift radio
galaxies, which is thought to be indicative of non-uniform
illumination consistent with the Unified Model of AGN.  It has been
inferred that the QSO and ULIRG phenomena are closely related, with
ULIRGs being postulated as being Type 2 objects, or ``QSO-2s''
\cite{Hines98}.  Genzel \etal (1998) \nocite{Genzel98} have made \ISO
observations of ULIRGs in order to determine which is the dominant
emission mechanism at FIR wavelengths.  They find that $\sim 75$ per
cent are powered by star-formation, while the remaining $\sim 25$ per
cent are AGN (see also \nocite{Rigopoulou96} Rigopoulou, Lawrence \&
Rowan-Robinson 1996).  A massive dusty torus will also contain huge
amounts of molecular gas, and therefore by definition will also be a
prime site for star-formation.  It has become increasingly evident
that both AGN and starburst activity are often present in the same
object, and therefore it is na\"{\i}ve to assume that the two
processes are either independent or mutually exclusive.

The first spectroscopic follow-up of the faint sub-mm sources tends to
show a mixture of dusty starburst and AGN.  However, in obscured
sources, it is frequently difficult spectroscopically to disentangle
starburst from AGN components.  Thus, in a SCUBA survey of four
distant clusters of galaxies \nocite{SIB97}(Smail, Ivison \& Blain
1997), the brightest source, SMM\,02399-0136, was found to be a
hyperluminous, active galaxy at redshift $z=2.8$ \cite{Ivison98}.
\citeANP{Frayer98} (1998) compare the measured $L_{\rm FIR}/L'_{\rm
CO}$ ratio of this high redshift galaxy, SMM\,02399-0136
\cite{Ivison98}, to that of the local starburst, Arp\,220, and find
that it is twice as high.  Since the FIR emission from Arp\,220 is
purely from merger-induced star-formation, they infer that
approximately half of the FIR emission from SMM\,02399-0136 must
therefore be due to a dust-enshrouded AGN.  Further examples include
the ultraluminous BALQSO, APM\,08279+5255, which at $z \sim 3.9$ is
apparently the most luminous object currently known \cite{Lewis98},
and the sub-mm source B\,1933+503, thought to be a high redshift
($z>2$) dusty radio quasar \cite{Chapman99}.  Such surveys suggest
that starburst galaxies dominate over AGN at faint sub-mm fluxes, but
the area of sky and number of sources detected is still quite small.

Of course, large populations of such dust-enshrouded QSOs and AGN have
been hypothesized in order to explain the observed shape of the X-ray
background (XRB) spectrum (Madau, Ghisellini \& Fabian
1994\nocite{M94}; \citeNP{C95}; \citeNP{KFG99}).  In addition, the
hard X-ray source counts are considerably higher than at soft X-ray
energies \cite{ASCA1}, implying the presence of an additional
hard-spectrum population to the broad-line QSOs known to make up the
bulk of the sources at lower energies.  However, one of the central
features of obscured QSO models is the presence of significant
quantities of gas and dust surrounding the active nucleus, providing
the means of absorbing a large fraction of the intrinsic radiation and
affecting the observed properties in the optical and X-ray regimes.
But this energy must escape somewhere, and through heating of the
dust, the absorbed flux is re-radiated at far infra-red (FIR)
wavelengths, providing an important test of the models.  The amounts
of dust invoked are quite considerable, and combined with the huge
quantity of energy that must be radiated, plus the wide redshift
distribution of these sources, would imply a substantial contribution
to the FIR/sub-mm background radiation and source counts.  This
motivates our investigation here into the impact of obscured QSO
models, used initially to explain observations at X-ray energies, at
much longer wavelengths.

While this paper was in preparation, \nocite{Almaini99} Almaini,
Lawrence \& Boyle (1999) have published sub-mm count predictions for
obscured QSOs which give similar results to those described here.
However, the conclusions of these authors are based on a simpler model
where both the fit of the obscured QSOs to the XRB and the obscured
QSO sub-mm spectrum have to be assumed.  We believe that the sub-mm
predictions presented here are more robust since they are based on
models which have been shown to give good fits to the X-ray background
over its full energy range.  In addition, in our method the absorption
of the X-ray and optical spectra to produce the sub-mm spectra leaves
no question as to their self-consistency.

In Section \ref{sec:submmobs}, we describe the latest measurements
of the spectrum and intensity of the far infra-red background, and
recent observations of the sub-mm source counts.  Section
\ref{sec:smmod} discusses how our obscured AGN model can be extended
to these longer wavelengths, and describes how the emission from the
obscuring dust torus has been modelled.  The results are presented in
Section \ref{sec:smresults}, and are compared with the observed sub-mm
number counts and the intensity of the far infra-red background at
these wavelengths.  We discuss the implications of our results in
Section \ref{sec:smdisc}, and make predictions for the number-redshift
relation for future $850\um$ surveys in Section \ref{sec:smfuture}.

Some of these results have already been discussed by
\citeANP{KFG99PhD} (1999).

\section{Far Infra-red and Sub-mm Observations}
\label{sec:submmobs}

It is extremely difficult to measure the extragalactic far infra-red
background (FIRB), due to the presence of foreground components from
interplanetary zodiacal dust emission (peaking at $\sim 20\um$) and
interstellar dust emission from our Galaxy (peaking at $\sim 150\um$).
This zodiacal and galactic contamination must be carefully modelled,
and subtracted to leave the extragalactic background.  In addition,
the cosmic microwave background (CMB) also contributes at these
wavelengths, peaking at $\sim 1\mm$, and this must also be accounted
for.  Once these components have been accurately modelled and removed,
it should then be possible to detect any extragalactic FIRB.

The first detection of the FIRB was claimed by \citeANP{Puget96}
(1996), who used data from {\em FIRAS} on board {\em COBE}, taking
advantage of the FIR window from $200 - 800\um$, between the peaks of
the zodiacal emission and the CMB, and found that the intensity of the
extragalactic background has the form:

\[
\nu I_\nu \simeq 3.4 \left(\frac{\lambda}{400\um}\right)^{-3} 
\times10^{-9} \fluxw, 
\]

\ni in the range $400\um < \lambda < 1000\um$.  More recent
measurements have been made at shorter wavelengths using {\em DIRBE},
also on board {\em COBE} (Schlegel, Finkbeiner \& Davis
1998\nocite{Schlegel98}; \citeNP{Hauser98}; \citeNP{Fixsen98}).

At longer wavelengths, there are several surveys currently in progress
with the aim of resolving the source populations contributing to the
sub-mm background.  These surveys take advantage of the increased
sensitivity and resolution now available with the SCUBA camera on the
JCMT.

SCUBA sub-mm surveys can be divided into two types, the first being
pointed observations of blank fields (\citeANP{Hughes98} 1998;
\citeNP{Barger98}; \citeNP{Eales99}).  The second strategy is to make
pointed observations of clusters of galaxies, in order to take
advantage of the gravitational amplification due to the lensing mass
of the cluster \nocite{SIB97} (Smail, Ivison \& Blain 1997).  In this
way, sources can be detected that would be fainter than the flux limit
possible without the amplification factor, and in addition, due to the
magnification of the source plane, there are fewer problems with
source confusion.  The $850\um$ number counts currently published can
be fitted with a power-law of $N(>S) = 7.9\times10^3 S^{-1.1}$, where
$N$ is the number of sources per square degree detected above a flux
limit of $S \mJy$ \cite{SIBK98}.  

These recent measurements of both the spectrum of the far infra-red
background, and the sub-mm number counts, now provide strict
constraints with which to test theories of galaxy evolution,
star-formation history, etc.  In the SCUBA Lens Survey \cite{SIBK98}
sample of sub-mm sources, at least 20 per cent were found to show
evidence for AGN activity \cite{Barger99}.  In the next Section, we
investigate the implications of the obscured AGN hypothesis, by
modelling their properties at sub-mm wavelengths.  In Section
\ref{sec:smresults}, we predict the number of such sources expected
from our models compared with the data described above, and put limits
on their contribution to the far infra-red background.

\section{Modelling}
\label{sec:smmod}

\begin{table*}
\caption{Parameters of four Boyle \etal (1994) model fits to the {\em
ROSAT} and {\em Einstein} EMSS QSO X-ray luminosity function and its
evolution, as used in our calculations.  The first column contains our
adopted descriptive model name, POW($q_0$) or POL($q_0$), with the
corresponding model name from Boyle \etal (1994) in the column headed
B94.  The XLF is parametrized by a broken power-law, with indices
$\gamma_1$ and $\gamma_2$ below and above the break luminosity
$L_X^*$.  The evolution is described by either power-law or polynomial
evolution, in both $q_0=0.0$ and $q_0=0.5$ universes.}
\begin{center}
\begin{tabular}{lcclccccccc} \hline
Model & B94 & $q_0$ & Evolution & $\gamma_1$ & $\gamma_2$ & 
$\log L^*_X(0)^\dag$ & $\gamma_z$ & $z_{{\it cut}}$ & 
$\gamma_z^{\prime}$ & $\Phi^*_X$$^\ddag$ 
\\ \hline
POW(0.0) & G & 0.0 & $(1+z)^{\gamma_z}+z_{{\it cut}}$ &
1.53 & 3.38 & 43.70 & 3.03 & 1.89 &    & 0.79 \\
POW(0.5) & H & 0.5 & $(1+z)^{\gamma_z}+z_{{\it cut}}$ &
1.36 & 3.37 & 43.57 & 2.90 & 1.73 &    & 1.45 \\
POL(0.0) & K & 0.0 & $\gamma_zz + \gamma_z^{\prime}z^2$ &
1.50 & 3.35 & 43.71 & 1.14 &    & -0.23 & 0.84 \\
POL(0.5) & L & 0.5 & $\gamma_zz + \gamma_z^{\prime}z^2$ &
1.26 & 3.32 & 43.52 & 1.15 &    & -0.26 & 1.89 \\ \hline 
\end{tabular}
\end{center}
\vspace*{4mm}
{\small $^\dag$where $L^*_X(0)$ is the $0.3-3.5\keV$ luminosity in
units of \ergps.}\\ {\small $^\ddag$where $\Phi^*_X$ is in units of
$10^{-6}\Mpc^{-3}(10^{44}\ergps)^{-1}$.}
\label{tab:models}
\end{table*}

\label{sec:moddescr}

Here we aim to use the obscured AGN models developed in
\citeANP{KFG99} (1999) in order to make self-consistent predictions at
sub-mm wavelengths.  We assume that the luminosity absorbed at high
energies is then reprocessed and emitted at lower energies, with a
known thermal spectrum.  The fluxes expected from such objects can
then be estimated, from which we can predict sub-mm source counts and
their contribution to the far infra-red background.  The assumptions
we have made are the following:

\begin{itemize}

\item The intrinsic $0.3-3.5\keV$ X-ray luminosity of each source is
known, and is assumed to come from the zero-redshift X-ray luminosity
function, with parameters as described in Table \ref{tab:models}.

\item The column density, $N_H$, perceived by the
optical, ultra-violet, and X-radiation is assumed to be a measure of
the intrinsic amount of dust present, and is not affected
significantly by the viewing angle.  The covering factor, $f_{cov}$,
of this obscuring material is defined to be the fraction of lines of
sight for which a constant $N_H$ is seen, and the remaining lines of
sight are unobscured.

\item All the absorbed flux goes into heating up the dust and gas in
the obscuring medium, whatever the assumed geometry, which is then
re-radiated isotropically in the thermal infra-red.

\end{itemize}

\subsection{Column density distribution}
\label{sec:tilt}

Several different column density distributions have been proposed in
order to explain the observed spectrum of the X-ray background.  For
example, Madau \etal (1994) used a bimodal model, consisting of a
population of unabsorbed sources plus 2.5 times as many absorbed
sources whose column densities had a Gaussian distribution with mean
$\log (N_H/\col) = 24.0 \pm 0.8$.  \citeANP{C95} (1995) used four
sub-classes of absorbed populations, each with a different
normalization with respect to the unabsorbed population.

\bfig
\psfig{figure=popsalt.ps,width=3.0in,angle=270}
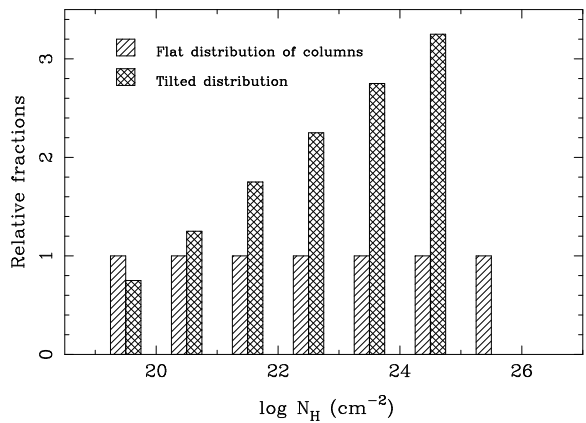
\caption{Comparison of the two distributions of obscured QSO
populations used in this paper, where the size of each population is
normalized relative to the unobscured $\log (N_H/\col) \sim 19.5$
population.  Our model of a flat distribution of columns is shown as
hatched regions, and the tilted distribution is shown as cross-hatched
regions, where the two models are shown slightly offset from one
another for clarity.  The tilted distribution has a greater proportion
of high column density objects, as used to improve the fits to the
data in a $q_0=0.5$ universe (Gunn \& Shanks 1999).}
\label{fig:model_distn_1}
\efig

Here we use two models, the first being a {\em flat} distribution of
columns, which was found in \citeANP{KFG99} (1999) to give good
agreement with the observed X-ray source counts and XRB spectrum for a
$q_0=0.0$ universe.  The second model has a {\em tilted} distribution,
with a greater proportion of high columns than in the flat model,
which gives better agreement with the data in a $q_0=0.5$ universe.
The differences in the observed properties of each population of
sources are attributable solely to the column density.  The {\em flat}
distribution uses seven populations evenly spaced in log space: $\log
(N_H/\col) = 19.5$, $20.5, 21.5, 22.5$, $23.5, 24.5, 25.5$, each
containing the same number and intrinsic luminosity function of AGN.
The {\em tilted} distribution uses six populations, with the relative
normalizations obeying the relation:

\[
\Phi (N_H) = \left\{1 + 0.5 \log \left( \frac{N_H}{10^{20}} \right) 
             \right\} \Phi^*.
\]

\ni The two distributions are compared in Fig.~\ref{fig:model_distn_1}.

\subsection{X-ray luminosity function and evolution}

Since we are assuming that the observed sub-mm emission results from
reprocessed nuclear X-ray and optical radiation, we take as a starting
point a two power-law zero-redshift X-ray luminosity function (XLF):

\[ 
\Phi_X (L_X) = \left\{ \begin{array}{ll} 
               \Phi^*_X L^{-\gamma1}_{X_{44}} 
                 & L_X < L^*_X(z=0) \\ 
               \Phi^*_X L^{-\gamma2}_{X_{44}}
                 L^{*(\gamma2 - \gamma1)}_{X_{44}}
                 & L_X > L^*_X(z=0),
               \end{array} \right.
\]

\ni where $\Phi^*$ is the normalization of the XLF, and $L_{X_{44}}$
is the $0.3 - 3.5 \keV$ X-ray luminosity in units of $10^{44} \ergps$.
Here the luminosity evolution is parametrized as either polynomial
evolution:

\[
L^*_X(z) = L^*_X(0) \, 10^{(\gamma_z z + \gamma_z^{\prime} z^2)},
\]

\ni or power-law evolution:
 
\[
L^*_X(z) = L^*_X(0) (1 + z)^{\gamma_z},
\]

\ni where a maximum redshift, $z_{{\it cut}}$, at which the evolution
stops is incorporated, such that: 
 
\[
L^*_X(z) = L^*_X(z_{{\it cut}}) \hspace{2cm} z > z_{{\it cut}}.
\]

\ni The parameters for the XLF and its evolution have been taken from
the fits to {\em ROSAT} and {\em Einstein} EMSS QSO number counts by
Boyle \etal (1994), and are listed in Table \ref{tab:models}.  The
broad line QSOs used to define the XLF correspond to our populations
with $\log (N_H/\col) = 19.5 {\rm ~and~} 20.5$, so we normalize the
populations accordingly.

\subsection{Canonical X-ray/optical QSO spectrum}

The intrinsic QSO spectrum, $F(E)$, is assumed here to consist of two
power-laws, with spectral indices $\alpha_x=0.9$ \cite{NP94} and
$\alpha_{opt}=0.8$ \cite{Francis93} at X-ray and optical energies
respectively.  In addition, the X-ray power-law is modified by the
effects of reflection.  The relative normalization is defined by a
power-law of slope $\alpha_{ox}=1.5$ joining $2\keV$ and $2500\Ang$
(\citeNP{Tanan79}; \citeANP{Yuan98_submmpaper} 1998).  The behaviour
of the spectrum between these two regimes is not well known however.
\citeANP{Zheng97} (1997) investigate the far ultra-violet properties
of a sample of high redshift QSOs and find that the radio-quiet QSO
spectrum can be approximated by a broken power-law, where the spectral
break occurs at $\sim 1050 \Ang$ ($\sim 0.01\keV$).  The spectral
index at longer wavelengths is $\alpha_{opt} \sim 0.86 \pm 0.01$ ({\em
c.f.}  $\alpha_{opt} \sim 0.8$ used here), steepening significantly to
shorter wavelengths (see also \citeNP{Laor97}).  For simplicity, we
therefore assume that the spectral break can be approximated by a step
discontinuity at $0.01\keV$.

\subsection{Absorbed X-ray/optical QSO spectrum}

The next step is to estimate the absorbed luminosity, $L_{\rm abs}$,
responsible for the heating of the torus, for each column density
used.  At X-ray energies, the opacity is dominated by photo-electric
absorption at columns of $N_H < 10^{24} \col$.  Above $N_H \sim
10^{24} \col$, the obscuring medium becomes Compton thick due to
electron scattering, such that the effective optical depth is
$\tau_{\rm eff} = \tau_{\rm ph} + \tau_{\rm es}$.  Since the absorbed
fraction of the luminosity is almost unity above $N_H \sim 10^{24}
\col$ due to photo-electric absorption alone (see
Fig.~\ref{fig:fracabs}), we do not take electron scattering into
account here as it does not affect our results significantly.  The
photo-electric absorption coefficients \cite{MM83} and can be
evaluated using {\sc xspec} \cite{Arnaud96}.  In the optical, the dust
extinction laws from \citeANP{Howarth83} (1983) and \citeANP{Seaton79}
(1979) are used.  A constant gas to dust ratio is implicitly assumed
here.  As the column increases, so does the amount of the continuum
which is destroyed by these processes.

The absorbed luminosity can be approximated by calculating the energy
at which the optical depth is unity for each process, $E(\tau_x=1)$
and $E(tau_{opt}=1)$ respectively, as a function of the column
density, and then by assuming that all the luminosity emitted between
these two energies is absorbed.  Using the relationship $\tau(E) =
\sigma(E) N_H$, the cross-section scales as $\sigma(E) = 1/N_H$, for
an optical depth of unity.  The absorbed luminosity is then calculated
from:

\[
L_{\rm abs} = f_{cov} 
  \int^{E(\tau_x = 1)}_{E(\tau_{opt} = 1)} F(E) \, \dd E.
\]

The range of intrinsic luminosities is defined by the X-ray luminosity
function (XLF), so the absorbed luminosity is normalized by the known
X-ray luminosity, $L_X$:

\[
L_X = \int^{3.5\keV}_{0.3\keV} F(E) \dd E.
\]

\ni The fraction of the luminosity which is absorbed can then be
calculated.  Fig.~\ref{fig:fracabs} shows how the absorbed fraction
increases with column density, saturating above $N_H \sim 10^{25}
\col$, as no radiation whatsoever can escape unaffected from the
nucleus, and all the emitted energy goes into heating up the obscuring
medium.

\bfig
\psfig{figure=fracabs.ps,width=3.0in,angle=270}
\caption
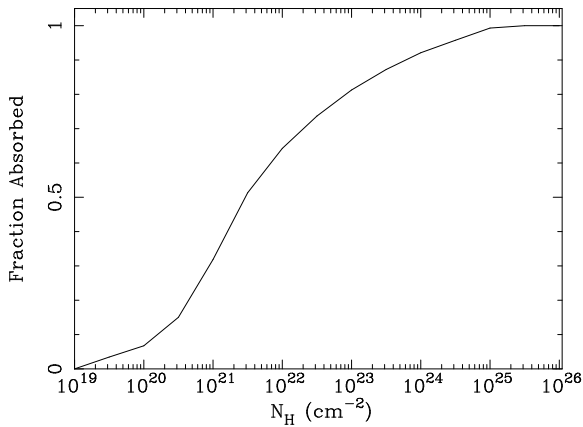
{The fraction of the total X-ray and optical luminosity absorbed, as a
function of the column density, $N_H$, for a covering factor of unity.
Once the column is as high as $N_H \sim 10^{25} \col$, effectively all
the energy is absorbed.  For $f_{cov}=0.5$, the fraction absorbed will
be half that shown here.}
\label{fig:fracabs}
\efig

\subsection{Far infra-red QSO spectrum}

Finally, we assume that all the absorbed radiation, $L_{\rm abs}$, has
to escape as thermal emission from the dust in the far infra-red, and
therefore that $L_{\rm FIR} = L_{\rm abs}$.  We also assume that the
FIR luminosity is isotropic, and therefore that the received flux is
independent of viewing angle.  For the case for which the obscuring
medium is also isotropic, this is a realistic assumption, as
self-shielding of the inner regions means that only radiation from the
outer, cooler layers of dust will be received.  However, for a
toroidal geometry (\eg, $f_{cov} \sim 0.5$), radiation from the hot
dust in the innermost regions will be able to escape from the top of
the torus, and for face-on viewing angles, the spectrum will be
broadened due to the superposition of components at a range of
different dust temperatures \cite{PK92b}.  This has the effect of
boosting the flux at short wavelengths ($\approxlt 10 \um$) but does
not change the spectrum significantly at the wavelengths at which we
are primarily concerned ($\approxgt 100 \um$), as the hot emission
component makes little contribution, and the torus is optically thick.
As the true geometry is still unclear, neglecting this effect means
that the predictions from our models at short wavelengths should be
taken to be lower limits.  We therefore choose to approximate our
obscuring medium by isothermal dust, emitting isotropically.

By convention, at these energies, frequency units are used, and we
write:

\[
L_{\rm FIR} = \int P(\nu_e) \, \dd \nu_e.
\]

Assuming optically thin dust emission, the Planck function, $B(\nu,
T)$, is modified by an opacity law, where the opacity depends on both
the dust grain composition and the size and shape distribution of the
grains, and which can be parametrized as $\kappa_d \propto \nu^\beta$.
Following \citeANP{Cimatti97} (1997), we use the opacity law:

\[
\kappa_d = 0.15 \left(\frac{\nu_e}{250 \, {\rm GHz}}\right)^2 \cm^2
\g^{-1}.
\]

\ni The dust temperature is taken to be in the range $30\K < T_d <
70\K$ (\citeNP{Haas98}; \citeNP{Benford98}, 1999\nocite{Benford99}).  
Since the emitted FIR luminosity is constrained by the X-ray
luminosity, the effect of increasing the assumed temperature means
that the total normalization must be reduced in order to keep $L_{\rm
FIR}$ constant, and vice versa.

The emitted power, $P(\nu_e)$, can therefore be parametrized as follows:

\be
P(\nu_e)=4\pi \kappa_d (\nu_e) B(\nu_e, T_d) M_d,
\label{eqn:pnue}
\ee

To calculate the received flux from such a source, we use the
relationship:

\[
S(\nu_o) \, \dd \nu_o = \frac{P(\nu_e)}{4\pi D_L^2} \, \dd \nu_e
              = (1 + z) \frac{P(\nu_e)}{4\pi D_L^2} \, \dd \nu_o,
\]

\ni where $D_L$ is the luminosity distance and $\dd \nu_e = (1 + z) \, 
\dd \nu_o$, to give:  

\be
S(\nu_o) = \frac{(1 + z) \kappa_d(\nu_e) B(\nu_e, T_d) M_d}{D_L^2}.
\ee

To calculate the integrated source counts as a function of flux
density, we then use the same method as in \citeANP{KFG99} (1999),
starting from the $0.3-3.5\keV$ XLF, and using the above relationships
between $L_X$ and $S(\nu_o)$.  Note that in the sub-mm regime,
traditionally flux {\em density} is used, with units of Janskys, ($1
\Jy = 10^{-23} \fluxerg \Hz^{-1}$), rather than the broad-band flux
used at X-ray energies (units: $\fluxerg$), or optical magnitudes.

\section{Sub-mm predictions}
\label{sec:smresults}

\begin{table*}
\caption[ ]
{The contribution to the intensity of the FIRB at $850\um$ predicted
by our obscured AGN model.  The parameters used are as described in
the text, where we have investigated the effects of changing the X-ray
luminosity function and evolution, the covering factor of the
obscuring medium $f_{cov}$ and its temperature $T_d$, and the maximum
redshift $z_{max}$.  The predictions are compared with the observed
$850\um$ background intensity from \citeANP{Fixsen98} (1998) of
$I_{\rm FIRB} = 5.03 \times 10^{-10}\fluxw$.}
\begin{center}
\begin{tabular}{lccccc}\hline
Model & $f_{cov}$ & $z_{max}$ & $T_d$ & $I_Q$ & $I_Q/I_{{\rm FIRB}}$  \\
      &           &          & ($\K$) & ($10^{-11}\fluxw$) & (\%) \\
\hline
POW(0.0)  & 1.0 & $5$  & $30$ & $8.70$  & $17.4$ \\
POW(0.5)  & 1.0 & $5$  & $30$ & $3.80$  & $ 7.6$ \\
POL(0.0)  & 1.0 & $5$  & $30$ & $5.41$  & $10.8$ \\
POL(0.5)  & 1.0 & $5$  & $30$ & $2.23$  & $ 4.5$ \\
POW(0.0)  & 0.5 & $5$  & $30$ & $6.85$  & $13.2$ \\
POW(0.5)  & 0.5 & $5$  & $30$ & $3.00$  & $ 6.0$ \\
POL(0.0)  & 0.5 & $5$  & $30$ & $4.08$  & $ 8.2$ \\
POL(0.5)  & 0.5 & $5$  & $30$ & $1.75$  & $ 3.5$ \\
POW(0.0)  & 1.0 & $2$  & $30$ & $1.30$  & $ 2.6$ \\
POW(0.0)  & 1.0 & $10$ & $30$ & $16.4$  & $32.9$ \\
POW(0.0)  & 1.0 & $5$  & $50$ & $1.04$  & $ 2.1$ \\
POW(0.0)  & 1.0 & $5$  & $70$ & $0.149$ & $ 0.3$ \\
POW(0.5)t & 1.0 & $5$  & $30$ & $7.57$  & $15.1$ \\
POL(0.0)t & 1.0 & $5$  & $30$ & $4.45$  & $ 8.9$ \\
POW(0.5)t & 0.5 & $5$  & $30$ & $5.24$  & $10.5$ \\
POL(0.0)t & 0.5 & $5$  & $30$ & $3.05$  & $ 6.1$ \\
\hline
\end{tabular}
\end{center}
\label{tab:firbint}
\end{table*}

Here we investigate the contribution to the sub-mm source counts and
the far infra-red background predicted by our obscured AGN model.  We
look at the effects of changing certain parameters, such as the
covering factor and temperature of the obscuring medium, the
luminosity evolution of the AGN, and the maximum redshift at which
these sources exist.  At present, these parameters are not well
constrained, particularly at high redshift.  However, despite evidence
that the space density of QSOs declines beyond $z\sim3$
\cite{Shaver96}, we know that QSOs exist at high redshift.  New search
techniques are discovering more such objects all the time, for
instance the three QSOs found recently by the SDSS Collaboration
\nocite{hizqso} (Fan \etal 1998), all with redshifts in the range
$4.75 < z < 5.0$.  As we are using luminosity functions and
evolutionary models determined from X-ray selected QSOs, these have
diverging properties above $z\sim2$, and these can therefore be taken
to span the range of likely properties.  By invoking the most extreme
cases, we are able to put firm upper limits on the obscured AGN
contribution to both the source counts and FIRB intensity.

\subsection{The covering factor of the absorbing material}

\bfig
\psfig{figure=fcov.ps,width=3.0in,angle=270}
\caption
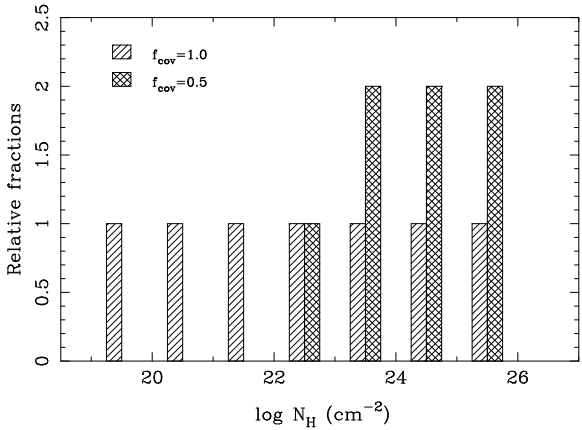
{Comparisons of two different intrinsic column density distributions
of obscured QSO populations, which give rise to the same {\em
perceived} flat distribution, as used in the $q0=0.5$ model.  For an
isotropic absorber, the covering factor is unity, \ie, $f_{cov}=1$,
and therefore an intrinsic flat distribution will also be perceived as
flat (hatched regions); however, for an absorber for which
$f_{cov}=0.5$, a larger number of obscured sources are required, as
half will be observed to be unobscured (cross-hatched regions); the
two distributions are shown slightly offset from one another for
clarity.  For the tilted distribution of columns used in $q_0=0.5$
scenarios, a similar correction must be made when $f_{cov}=0.5$ is
assumed.}
\label{fig:fcov}
\efig

In order to remain consistent with the assumed flat column
distribution used in \citeANP{KFG99} (1999), we must also consider the
covering factor of the obscuring material, as this affects the
intrinsic distribution of column densities.  Here, we take two
examples, first that the obscuring material is isotropic, \ie, with a
covering factor $f_{cov}=1$, and secondly that a torus covers half the
sky as perceived by the nucleus for all sources, \ie, $f_{cov}=0.5$.
We assume that these two values will span the true range of covering
factors.  Two intrinsic distributions giving rise to a {\em perceived}
flat column distribution, as used in the $q0=0.5$ model, are shown in
Fig.~\ref{fig:fcov}.

A method for differentiating between these two proposed geometries for
the obscuring material in AGN, is to determine the fraction which are
highly luminous at sub-mm wavelengths.  By definition, very little
dust exists along the line-of sight to broad-line QSOs.  Therefore, if
the absorbing medium is isotropic, the dust content of broad-line QSOs
must be intrinsically low, with low sub-mm emission.  However, for a
toroidal structure in the spirit of the Unified Model
\cite{Antonucci93}, broad-line QSOs could contain large amounts of
dust in a plane perpendicular to the line of sight.  This is
consistent with the large infra-red bump observed in the $3000\Ang$ to
$300\um$ spectrum of a sample of Palomar Green QSOs \cite{Sanders89}.
Therefore, if all QSOs contain large quantities of dust, they will be
strong sub-mm sources.  Proposed sub-mm observations of X-ray selected
QSOs using SCUBA/JCMT will shed further light on this question in the
near future.

\subsection{The effects of changing the XLF parameters}

We first investigate the predicted source counts for the obscured QSO
model with a flat distribution of columns, integrated over redshifts
$0<z<5$.  The X-ray luminosity function used is evolved according to
the power-law and polynomial models from \citeANP{B94} (1994), for
both $q_0=0.0$ and $q_0=0.5$ cosmologies.  The parameters of these
models are detailed in Table \ref{tab:models}.

\bfig
\psfig{figure=smcounts.ps,width=3.0in,angle=270}
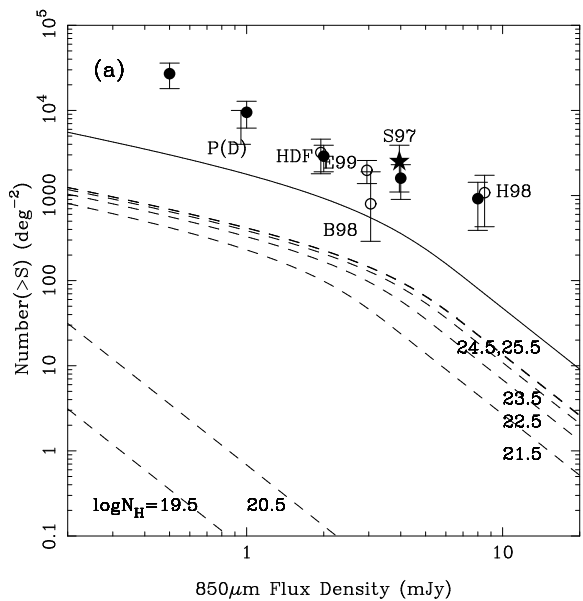
\caption{Predicted $850\um$ source counts of obscured QSOs, compared
with the observed counts as described in Section \ref{sec:submmobs}.
The filled circles show the results of the SCUBA Lens Survey (Blain
\etal 1999), and the other points are as labelled: S97 - Smail,
Ivison \& Blain (1997); B98 - Barger \etal (1998); H98 - Holland \etal
(1998); E99 - Eales \etal (1999); HDF, P(D) - Hughes \etal (1998).  A
dust temperature of $T_d = 30\K$ and a covering factor $f_{cov}=1$ is
assumed, integrated over $0<z<5$, for power-law luminosity evolution
and $q_0=0.0$, model POW(0.0).  The contribution from each individual
population of obscured QSOs is shown as a separate dashed line, marked
with the column density, and the total is shown by the solid line.  In
contrast to the X-ray number counts, here the biggest contribution
comes from the most highly obscured objects, since these contain the
largest amounts of dust.  This model accounts for $\sim 17$ per cent
of the FIRB.}
\label{fig:smcounts}
\efig
\nocite{Blain99}
\nocite{SIB97} 
\nocite{Barger98}
\nocite{Holland99}
\nocite{Eales99}
\nocite{Hughes98}

\bfigw
\begin{minipage}{3.2in}
\psfig{figure=smc_all_fc1.ps,width=3.0in,angle=270}
\end{minipage}
\hspace*{0.2in}
\begin{minipage}{3.2in}
\psfig{figure=smc_all_fc5.ps,width=3.0in,angle=270}
\end{minipage}
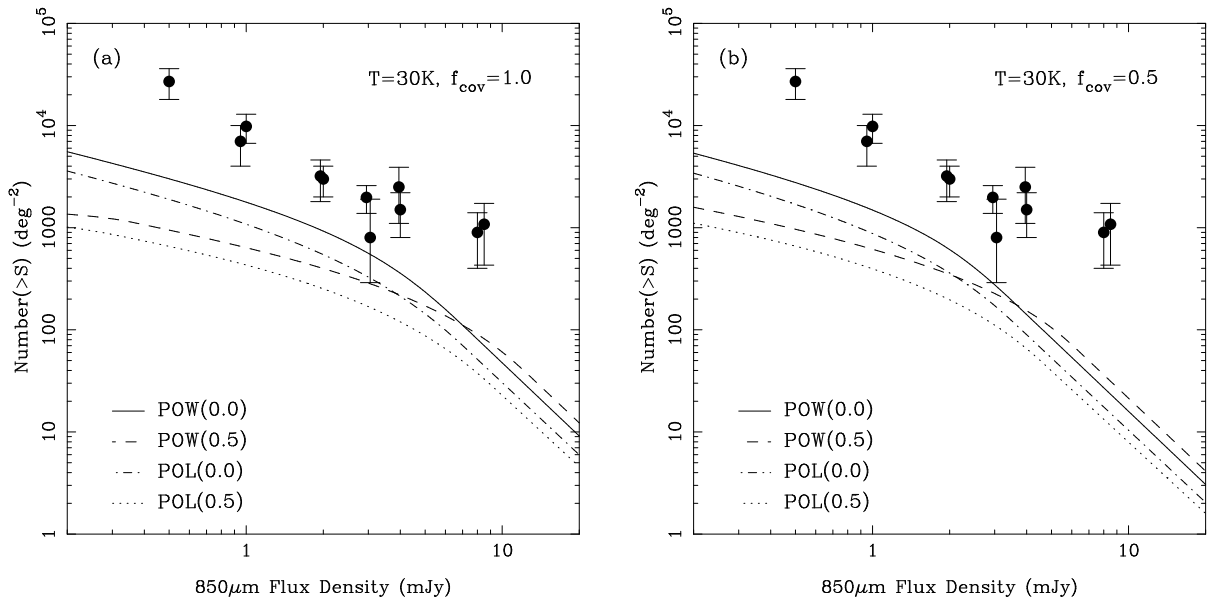
\caption{Total predicted $850\um$ source counts of obscured QSOs for
models using power-law and polynomial evolution, $q_0=0.0$ and
$q_0=0.5$ cosmologies, compared with the observed sub-mm counts as
described in Fig.~\ref{fig:smcounts}.  A covering factor of (a)
$f_{cov}=1$, and (b) $f_{cov}=0.5$, is assumed, with a dust
temperature of $T_d = 30\K$, integrated over $0<z<5$.  These models
can account for between $17$ per cent (POW(0.0), $f_{cov}=1$) and $4$ 
per cent (POL(0.5), $f_{cov}=0.5$) of the FIRB.  Note the change of
scale with respect to Fig.~\ref{fig:smcounts}.}
\label{fig:smcounts_fcov}
\efigw

First, we take as our fiducial model a power-law prescription for the
luminosity evolution, and a $q_0=0.0$ cosmology, which we denote as
POW(0.0).  This model will be used hereafter, unless stated otherwise.
In Fig.~\ref{fig:smcounts}, we show the source counts predicted by
this model, compared with the SCUBA data described in Section
\ref{sec:submmobs} and Fig.~\ref{fig:smcounts}.  A covering factor
of $f_{cov}=1$ is assumed, for a dust temperature of $T_d = 30\K$,
integrated over $0<z<5$.  The contribution from each population of
obscured QSOs is shown by a dashed line, with the total source counts
denoted by a solid line.  The reverse trend to that observed at X-ray
energies is seen here: the populations with the smallest column
densities give the lowest sub-mm fluxes, with the high column density
populations dominating the sub-mm source counts due to the large dust
masses present.  This model predicts $\sim 20$ per cent of the source
counts at $2\mJy$, and flattens off at fainter fluxes, but will
provide $\sim 17$ per cent of the FIRB at $850\um$.

In Fig.~\ref{fig:smcounts_fcov}, the predictions for all four Boyle
\etal models are shown, for covering factors of (a) $f_{cov}=1$ and
(b) $f_{cov}=0.5$, using the intrinsic column distribution that each
covering factor implies.  As expected, the low $q_0$ models give a
much larger contribution to the source counts than the $q_0=0.5$
models, due to volume effects.  Taking the measured value for the
$850\um$ FIRB intensity of $I_{\rm FIRB} = 5.0 \times 10^{-10}\fluxw$
from \citeANP{Fixsen98} (1998), we calculate the fraction of the FIRB
which can be accounted for by our obscured QSOs.  Low $q_0$ models
provide $\sim 8-17$ per cent of the FIRB intensity, compared with
$\sim 3-8$ per cent for the $q_0=0.5$ models, as shown in Table
\ref{tab:firbint}.  

The difference in the predicted number counts and background intensity
between the $f_{cov}=1$ and $f_{cov}=0.5$ cases is due to the
increased intrinsic far infra-red luminosity of sources which have an
isotropic absorbing medium, as a greater fraction of the nuclear
radiation is intercepted.  These intrinsically brighter sources are
therefore detected at higher fluxes, increasing the bright-end number
counts.  However, at faint fluxes, the predictions are similar for
both assumed covering factors, implying that we are observing high
redshift objects, which by definition will be the most luminous
sources in either case.

For an isotropic covering medium, the predicted source counts of
obscured AGN at $2\mJy$ account for between 8 and 20 per cent of the
total counts, whereas for the anisotropic case, the numbers drop by
about a third, since the objects are less luminous.  The contribution
to the intensity of the FIRB is $\sim 3-13$ per cent in 
the $f_{cov} = 0.5$ case, which is about $25$ per cent lower than 
the $\sim 4-17$ per cent predicted for $f_{cov}=1$.

\subsection{The effects of changing $z_{max}$ and $T_d$}

\bfig
\psfig{figure=smred.ps,width=3.0in,angle=270}
\vspace*{0.2in}
\psfig{figure=smtemp.ps,width=3.0in,angle=270}
\caption
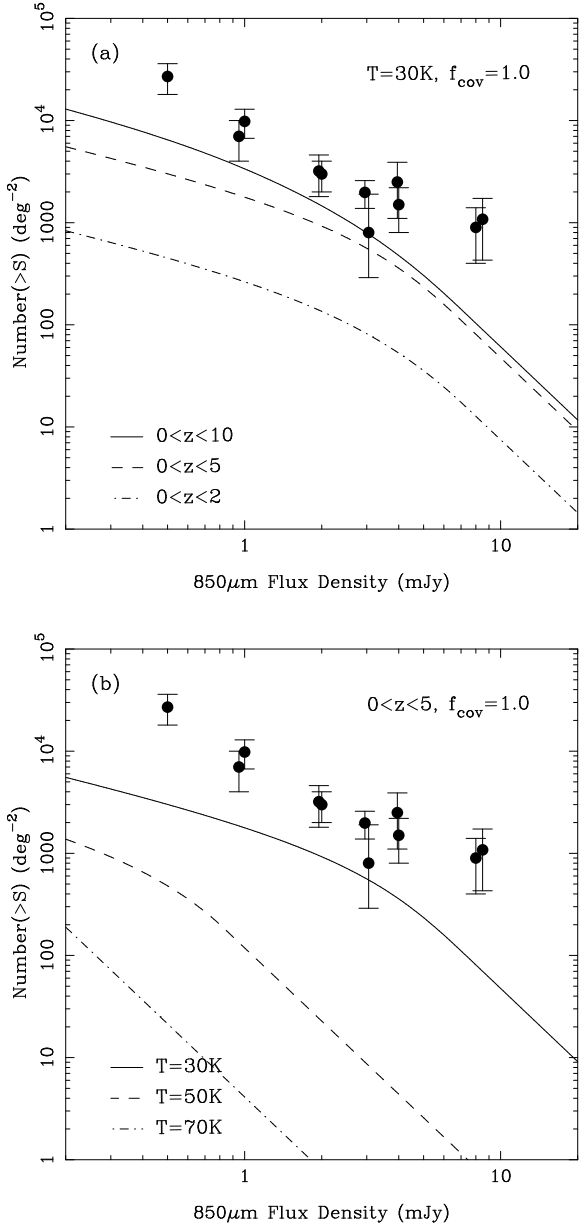
{(a) The effect on the observed $850\um$ source counts, of varying the
maximum redshift, $z_{max}$, for dust at $T_d=30\K$, for $z_{max}=2,5$
and $10$.  (b) The effect on the observed $850\um$ source counts, of
varying the dust temperature between $T_d=30\K$ and $T_d=70\K$, over
redshifts $0<z<5$.  In both cases, parameters for our fiducial model
POW(0.0) have been used, for seven populations with a flat column
distribution, and a covering factor $f_{cov}=1$.}
\label{fig:smredtemp}
\efig

If we were to assume that obscured QSOs exist out to redshift
$z_{max}\sim10$, then we can make a much larger contribution to the
source counts and FIRB intensity, as shown for $T_d=30\K$ and
$f_{cov}=1$ in Fig.~\ref{fig:smredtemp}(a).  However, it is difficult
to envisage a scenario in which large numbers of dusty QSOs form at
such early epochs \cite{ER88}, and therefore this places a useful
constraint on the maximum contribution such models can provide.
Obscured QSOs from $0<z<10$ could contribute over $30$ per cent of the
\citeANP{Fixsen98} (1998) measurement of $I_{\rm FIRB}$.

There are very few complete sub-mm surveys of AGN at low and high
redshifts in the literature, with enough data to constrain the dust
temperatures in such objects.  The uncertainties involved in making
such estimates are high, where assumptions have to be made about the
cosmology, the dust masses and the opacities, and in addition,
observations at several different wavelengths are required in order to
start to put limits on the parameters in the models.  Treating the
dust as isothermal is likely to be an oversimplification, but will be
adequate for our purposes, and therefore we estimate that the range
$30\K < T_d < 70\K$ should span the most probable temperatures.  

For each column density used, the FIR luminosity is in effect fixed by
the known amount of nuclear X-ray and optical luminosity which has
been absorbed.  If we increase the assumed dust temperature, the mass
of the dust must be decreased in order to keep $L_{\rm FIR}$ constant.
As the temperature is increased, the peak of the thermal radiation
moves to higher frequencies, thereby reducing the intensity of the
source at sub-mm energies.  In Fig.~\ref{fig:smredtemp}(b), we show
the effect of varying the temperature on the sub-mm counts, where the
highest number of sub-mm sources are observed for the model with the
lowest dust temperature.  If the dust around most AGN is warm, $T_d
\sim 70 \K$, then the contribution to the $850\um$ FIRB will be
negligible, whereas at $100\um$, their impact will be more
significant.

\subsection{Tilted column distributions for $q_0=0.5$ models}

\bfig
\psfig{figure=smc_tilt.ps,width=3.0in,angle=270}
\caption
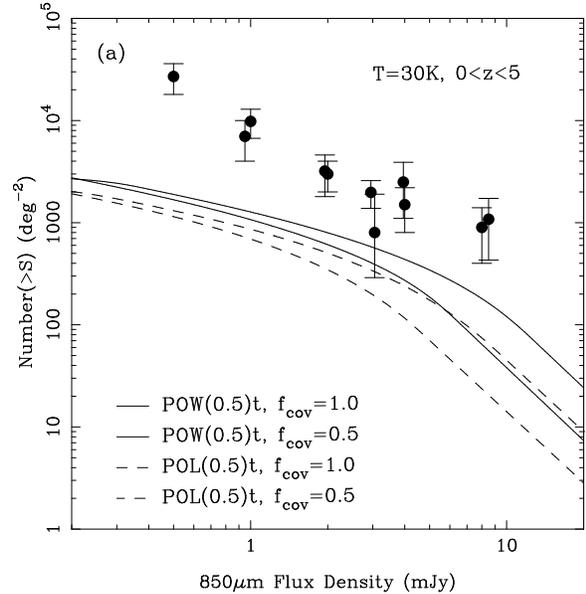
{The predicted total source counts for the $q_0=0.5$ models, for
power-law (solid lines) and polynomial (dashed lines) luminosity
evolution, using a tilted distribution of column densities.
Predictions for both a covering factor of $f_{cov}=1$ (bold lines) and
$f_{cov}=0.5$ (thin lines) have been plotted, using $T_d = 30\K$ and
$z_{max} = 5$.}
\label{fig:smtilt}
\efig

One further test here for the obscured QSO model is to look at the
tilted distribution of column densities, which was invoked in order to
obtain a better fit to the XRB for high density, $q_0=0.5$, models for
the XLF.  By skewing the distribution of objects towards higher column
densities, this could have the effect of overpredicting the source
counts and background intensity.  However, in Fig.~\ref{fig:smtilt},
it can be seen that the source counts predicted with this distribution
for a range of models lie well below the observed counts, and flatten
off towards fainter fluxes, again primarily due to volume effects.
The maximum contribution to $I_{\rm FIRB}$ from these models is $\sim
15$ per cent, for model POW(0.5)t with $T_d=30\K$ and $f_{cov}=1$.

\subsection{Number-redshift distributions}

\bfig
\psfig{figure=nztot_fc1_2mJy.ps,width=3.0in,angle=270}
\vspace*{0.2in}
\psfig{figure=nztot_fc5_2mJy.ps,width=3.0in,angle=270}
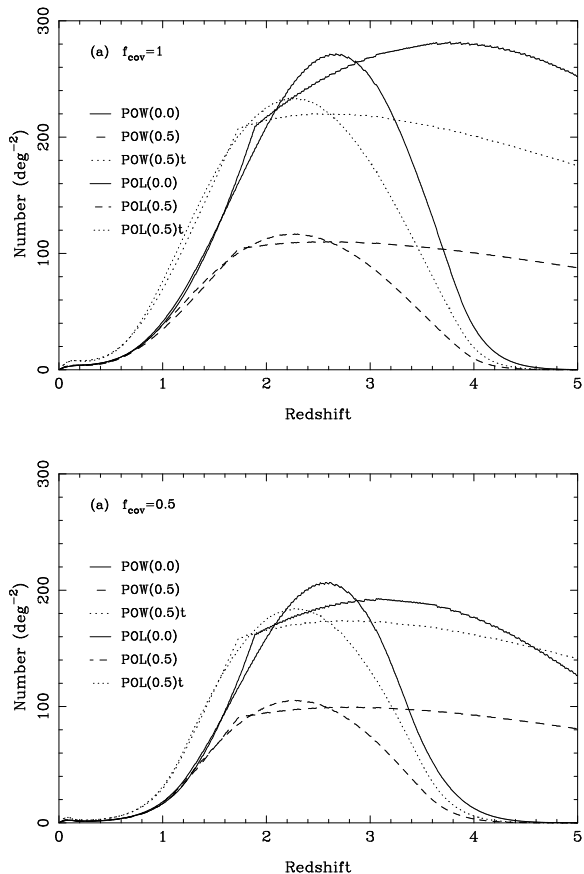
\caption{Predicted number-redshift distribution for an $850\um$ survey
to $2\mJy$, of obscured QSOS with $T_d=30\K$, and $z_{max}=5$, for
covering factors of (a) $f_{cov}=1$, and (b) $f_{cov}=0.5$.  Note the
higher number densities predicted for low $q_0$ models.}
\label{fig:smnz}
\efig

In Fig.~\ref{fig:smnz}, we plot the number-redshift distributions
predicted by our obscured QSO models, for an $850\um$ survey to
$2\mJy$ ({\em c.f.} the $1\sigma$ confusion limit for JCMT/SCUBA of
$0.44\mJy \, {\rm beam}^{-1}$; \nocite{Blain98} Blain, Ivison \& Smail
1998).  Again, we use a dust temperature of $T_d=30\K$, and consider
sources out to redshift $z_{max}=5$, and make predictions for both (a)
$f_{cov}=1$ and (b) $f_{cov}=0.5$.  We have taken four models with a
flat column distribution of seven populations of obscured AGN, and two
models using a tilted distribution of six populations, as described in
Section \ref{sec:tilt} (all for $f_{cov}=0.5$).  In the $q_0=0.0$
cosmology, the number of sub-mm sources predicted at redshifts $z>3$
is much higher than for $q_0=0.5$, consistent with the steeper
faint-end slope of the LogN:LogS relation for a low $q_0$ universe.
Once more source identifications become available, it will be very
straightforward to determine the form of the luminosity evolution, due
to the fact that it is as easy to detect a sub-mm source at high
redshift as at low redshift, and therefore if large numbers of high
redshift obscured QSOs exist, they will be found in deep SCUBA
surveys.

The follow-up identification programs for the latest SCUBA sub-mm
surveys are nearing completion, and the source catalogues will soon be
published.  We shall therefore be able to compare the numbers of AGN
detected with the redshift distributions predicted by our models.  It
is interesting to note that $\sim 10$ per cent of the optical
counterparts to sub-mm sources are classified as Extremely Red Objects
(EROs; \citeNP{Hu94}), which are found in \kband images of the SCUBA
error boxes \cite{Smail99}.  Deep radio maps provide more accurate
positional information than the SCUBA maps, assuming that the radio
and sub-mm emission is due to the same mechanism, and the ERO
counterparts are confirmed by the radio data.  It will be very
difficult to obtain spectroscopic redshift information about these
objects, even with a 10-m class telescope, as they are so faint in the
optical, with $I \gg 25$.  However, near infra-red spectroscopy has
been used successfully to obtain a redshift of $z=1.44$ for the ERO
HR10 \cite{Dey99}.  Potentially, these could be examples of very
highly obscured AGN at high redshift, for which in the optical and
near infra-red we see the dusty host galaxy (hence the very red
colours), and only see evidence for the AGN through the sub-mm and
radio emission.  Future deep surveys with {\em AXAF} and {\em XMM},
such as the proposed {\em AXAF} and SCUBA observations of the SSA\,13
field (Cowie \etalc), will provide a vital test of these theories, in
searching for faint X-ray emission associated with these sources.
However, unlike in the sub-mm regime, the X-ray $k$-correction is not
working in our favour, and if we assume that these sources are at high
redshift, then this is still an extremely ambitious project.

\subsection{Predicted spectrum of the far infra-red/sub-mm background}

\bfig
\psfig{figure=cibcomp.ps,width=3.0in,angle=270}
\caption
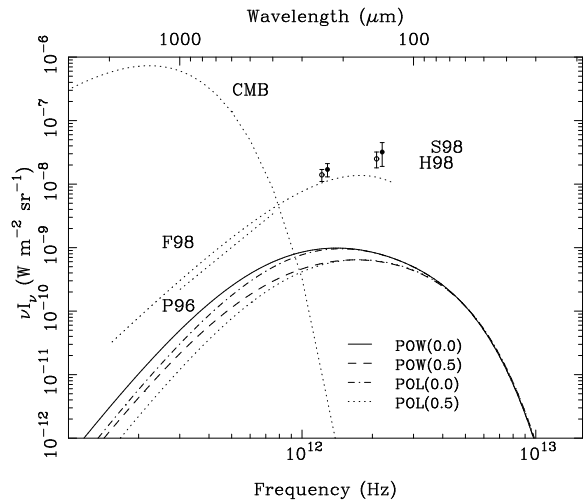
{The predicted contribution to the FIRB from seven populations of
obscured AGN with a flat column distribution, using power-law and
polynomial luminosity evolution, for both $q_0=0.0$ and $q_0=0.5$
cosmologies, as labelled.  A covering factor $f_{cov}=1$ is used, with
a dust temperature $T_d=30\K$, integrated over redshifts $0<z<5$.  The
data are as described in Section \ref{sec:submmobs}: CMB - Cosmic
Microwave Background, Mather \etal (1994); F98 - Fixsen
\etal (1998); P96 - Puget \etal (1996); H98 - Hauser \etal (1998); S98
- Schlegel \etal (1998).}
\label{fig:cibghkl}
\efig
\nocite{Mather94} 
\nocite{Fixsen98}
\nocite{Puget96}
\nocite{Hauser98}
\nocite{Schlegel98} 

We have shown from the sub-mm number counts predicted by our models
that obscured AGN provide a small but non-negligible fraction of the
intensity of the far infra-red background at $850\um$, ranging from 1
to 33 per cent.  Fig.~\ref{fig:cibghkl} shows the spectrum predicted
from our flat distribution of column densities using a covering factor
$f_{cov} = 1$, a dust temperature $T_d = 30\K$, and integrated over
redshifts $0<z<5$.  It can be seen that all four models predict the
same intensity at high frequencies.  This is to be expected, as the
emission here is dominated by low redshift objects for which the
rest-frame peak of the $30\K$ thermal spectrum is at $\sim 100\um$.
Since the model parameters have been obtained from fits of the X-ray
luminosity function to X-ray selected QSOs from {\em ROSAT} and the
{\em Einstein} EMSS, which have relatively low median redshifts of
$z\sim1.5$ and $z\sim0.2$ respectively, we would expect the
predictions to diverge at high redshift and therefore low frequencies.

\bfig
\psfig{figure=cibcomp_z.ps,width=3.0in,angle=270}
\vspace*{0.2in}
\psfig{figure=cibcomp_temp.ps,width=3.0in,angle=270}
\caption
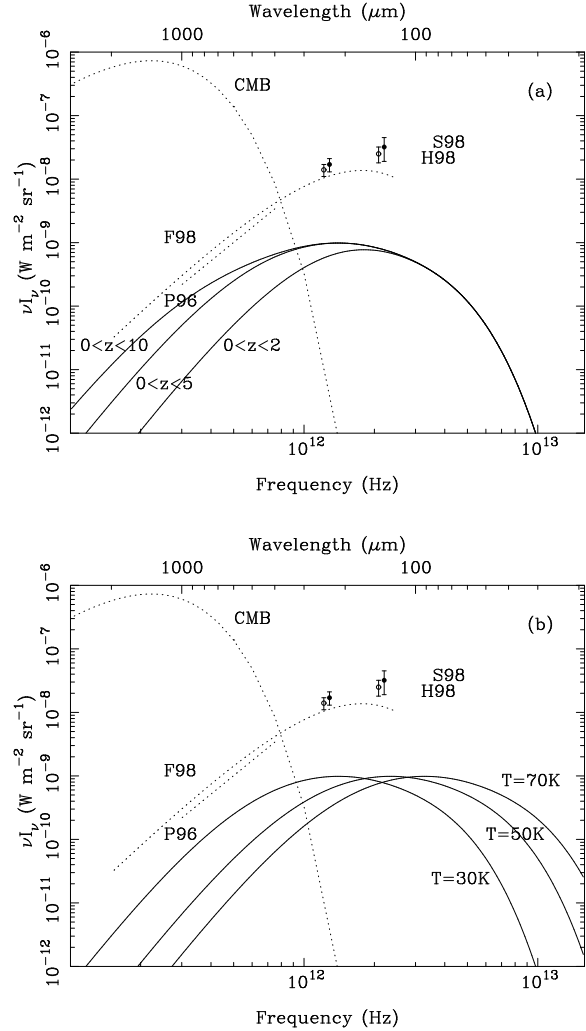
{(a) The effect on the predicted FIRB spectrum of varying the maximum
redshift, $z_{max}$, for dust at $T_d=30\K$, for $z_{max}=2,5$ and
$10$.  (b) The effect on the predicted FIRB spectrum of varying the
dust temperature between $T_d=30\K$ and $T_d=70\K$, over redshifts
$0<z<5$.  In both cases, our fiducial model has been used, for seven
populations with a flat column distribution, and a covering factor
$f_{cov}=1$.  The data are as described in Section \ref{sec:submmobs}.}
\label{fig:cibredtemp}
\efig

In Fig.~\ref{fig:cibredtemp}, we show the effects of changing the
maximum redshift and the dust temperature used in the models.  In
panel (a), it can be seen that the low redshift sources contribute at
higher frequencies around $100\um$, whereas the high redshift sources
account for the largest fraction of the FIRB intensity at $850\um$.
In panel (b), we show how as the dust temperature is increased, the
peak of the predicted FIRB background moves to higher energies.  The
integrated luminosity is fixed by the amount of absorbed nuclear X-ray
and optical radiation, and is independent of temperature.  Since this
is a spectral energy distribution diagram, in which equal areas mean
equal energies, then the predicted intensities for each model are
identical after a shift along the frequency axis.  A dust temperature
of $T_d \sim 40\K$ would give the maximum contribution to the peak of
the observed FIRB.

\section{Discussion}
\label{sec:smdisc}

The first question to address is what these obscured AGN will look
like at wavelengths other than the sub-mm.  Compared with the majority
of the X-ray source population, sub-mm sources have much higher dust
masses and therefore much higher column densities.  Hence in general,
the optical and X-ray nuclear emission will be totally obscured, and
the counterparts are likely to appear like relatively ``normal''
galaxies.  These galaxies may perhaps have narrow emission lines in
their optical spectra, possibly of high ionization species, or may
look very dusty from their optical and infra-red colours.  However, if
the obscuring material is not assumed to be isotropic, then a fraction
of these highly obscured sources will be orientated such that our line
of sight lies within the opening angle of the torus, and therefore the
X-ray and optical nuclear flux will escape unattenuated, while
simultaneously a large sub-mm flux is detected from the dust in the
torus.

The identification of optical counterparts to sub-mm sources has
similar problems to those encountered with X-ray data, in that the
point-spread function of the telescope is large (Half Power Beam Width
$= 14\arcsec\!\!.7$ at $850\um$ for SCUBA/JCMT), and therefore a
number of plausible counterparts can lie in the error box.  As sub-mm
sources often have associated radio emission, deep radio maps of
sub-mm survey fields have been taken (\citeANP{Ivison99} in
preparation; \citeANP{Richards99} 1998).  Star-formation regions are
expected to contain many supernova remnants, known to be strong radio
emitters, and even radio-quiet AGN are likely to be detected as very
faint, $\mu{\rm Jy}$ sources in extremely deep radio maps.  The high
angular resolution of the radio data combined with the fact that radio
emission is not affected by the presence of dust, means that optical
counterparts to the sub-mm sources can be found in a much less
ambiguous manner.

Once the counterpart has been found, the mechanism giving rise to the
sub-mm emission must then be determined.  By assuming that for
starburst galaxies, the star-formation rate controls both the radio
emission and the thermal sub-mm emission, \citeANP{Carilli99} (1999)
use the radio to sub-mm spectral index as a redshift indicator.  If
however, an independent measure of the source redshift can be
obtained, then the radio to sub-mm spectral index can be used to infer
whether there is any additional contribution to either component due
to the presence of an AGN.  A radio-loud AGN will have proportionally
higher radio emission than a starburst galaxy, whereas an obscured
radio-quiet AGN will have lower radio and higher sub-mm emission.

At present, there is little information available with which to
constrain the dust temperature or range of temperatures that should be
adopted for our models.  Here, we have assumed that $T_d$ remains
constant with redshift, but as shown in Fig.~\ref{fig:smredtemp},
the temperature affects the predicted source counts significantly.
The advent of larger samples of AGN at both low and high redshifts
with multi-wavelength FIR/sub-mm data will enable any evolution of the
dust temperature to be constrained.

Having chosen to keep $T_d$ constant with redshift, and a universal
opacity law, the only unconstrained parameter defining the luminosity
is therefore the dust mass, $M_d$, (see Eqn \ref{eqn:pnue}).
Therefore, since the FIR luminosity is directly related to the X-ray
luminosity, which we have modelled to undergo pure luminosity
evolution with redshift, then one possible interpretation is that the
dust mass also evolves with redshift, with the form $M_d \propto
(1+z)^3$.  If the dust mass does not scale with the luminosity, then
in order for our models to hold, there must be a balance between dust
mass and temperature, with the mass compensating for changes in
temperature.

A further consideration is the question of whether the assumption of
constant gas to dust ratio at all redshifts is appropriate.  For a
QSO, there exists a radius within which the radiation field is so
intense that dust particles will not survive, the so-called dust
sublimation radius.  \nocite{Granato97} Granato, Danese \&
Franceschini (1997) proposed that a significant quantity of gas exists
inside the dust sublimation radius, and that a large proportion of the
photo-electric absorption occurs within this region.  This would have
the effect of reducing the dust masses calculated using the column
densities inferred from the observed photo-electric absorption at
X-ray energies.  This would in turn, lower the sub-mm fluxes from
obscured AGN, cutting their contribution to the source counts and the
intensity of the FIR background.  However, the calculation of the gas
to dust ratio for SMM\,02399-0136 at redshift $z=2.8$ by
\citeANP{Frayer98} (1998), and SMM\,14011+0252 at redshift $z=2.6$ by
\citeANP{Frayer99} (1999), from measurements of CO line emission
combined with the sub-mm flux, gives a value similar to that found in
nearby galaxies, from which they infer that CO emitting sources at
high-redshift have already undergone significant chemical evolution.
We have therefore chosen to adopt a constant gas to dust ratio in our
models, since at present the available observations are insufficient
to constrain any suitable alternatives.

\section{Future observational tests of the models}
\label{sec:smfuture}

The next decade promises to bring enormous advances in this field,
with the advent of innovative instrumentation combined with the
light-gathering capacity of 10-m class telescopes, plus the new
generation of satellite-bourne detectors.  In the near infra-red,
large area surveys will be possible with new wide-field cameras, such
as the Cambridge Infra-Red Survey Instrument (CIRSI; \citeNP{CIRSI}),
allowing follow-up observations of the entire survey area for existing
and future X-ray surveys, rather than the pointed observations of
individual sources used previously.  In the far infra-red, the FIRBACK
survey \cite{FIRBACK} is an \ISOPHOT $175\um$ survey of $4 \deg^2$ of
sky at high Galactic latitudes, with the aim of determining the source
populations making up the far infra-red background. Proposed
satellite-borne mid and far infra-red observatories include the NASA
{\em Space Infra-Red Telescope Facility} ({\em SIRTF};
\citeNP{SIRTF}), and the ESA {\em Far Infra-Red and Sub-millimetre
Telescope} ({\em FIRST}; \citeNP{FIRSTa}; \citeNP{FIRSTb}) scheduled
for launch in 2001 and 2007 respectively.

\bfig
\psfig{figure=nztot_fc1_8mJy.ps,width=3.0in,angle=270}
\vspace*{0.2in}
\psfig{figure=nztot_fc5_8mJy.ps,width=3.0in,angle=270}
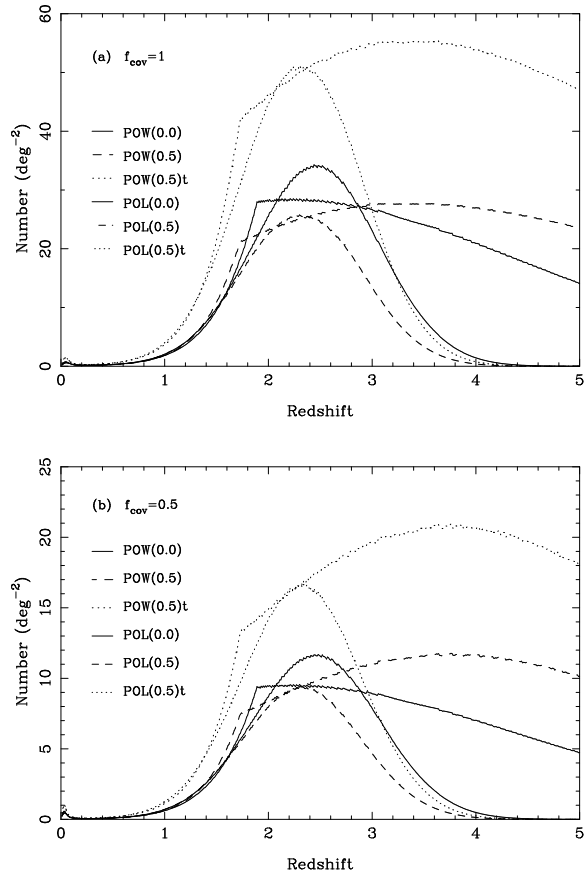
\caption{Predicted number-redshift distribution for an $850\um$ survey
to $8\mJy$, of obscured QSOS with $T_d=30\K$, and $z_{max}=5$, for (a)
$f_{cov}=1$ and (b) $f_{cov}=0.5$.  As in Fig.~\ref{fig:smnz}, we plot
the $q_0=0$ models (solid lines) and $q_0=0.5$ models (dashed lines)
for seven populations with a flat column distribution, and the tilted
$q_0=0.5$ models with six populations (dotted lines).  The power-law
models are plotted with bold lines in each case, and the polynomial
models are plotted with thin lines.}
\label{fig:smnzbr}
\efig

In the sub-mm regime, a proposed wide area survey with SCUBA plans
to cover a subset of the European Large Area \ISO Survey region (ELAIS;
\citeNP{Oliver98}) of around $640$ square arcminutes ($0.178 \,
\deg^2$).  The survey will have a brighter flux limit than existing
deep pencil-beam surveys, and aims for a $3\sigma$ detection threshold
of $\sim 8\mJy$, from which $\sim 40$ sources are expected.  Our
predictions for the number-redshift distribution for such a survey are
presented in Fig.~\ref{fig:smnzbr}(a) for $f_{cov}=1$, with the more
conservative predictions using $f_{cov}=0.5$ in
Fig.~\ref{fig:smnzbr}(b).  The expected number of sources ranges
between $\sim 14 \deg^{-2}$ (POL(0.5), $f_{cov}=0.5$, $\zmax=5$) and
$\sim 180 \deg^{-2}$ (POW(0.5)t, $f_{cov}=1$, $\zmax=5$), depending on
the evolution and column density distribution used in the model.
Clearly, if the obscuring torus is also a site of active
star-formation and the sub-mm flux of each component is comparable
\cite{Frayer98}, then the number of AGN may be somewhat higher than
predicted on the basis of the above model.

\section{Conclusions}
\label{sec:smconcl}

In this paper, we have extended the obscured QSO model for the X-ray
background of \citeANP{KFG99} (1999) to the sub-millimetre regime, by
considering the fate of the X-ray, ultra-violet and optical energy
absorbed by the obscuring medium.  This energy goes into heating up
the dust in the obscuring material, which then radiates thermally at
far infra-red and sub-millimetre wavelengths.  We have modelled the
obscuring medium as either isotropic or having a toroidal geometry,
which then dictates the intrinsic column density distributions which
are consistent with the line-of-sight distributions found from X-ray
and optical observations.

Since the spectrum in the sub-mm is rising steeply to higher
frequencies, the $k$-correction obtained is such that our obscured
QSOs are equally visible for redshifts of $1<z<10$.  We therefore use
the observed sub-mm source counts at $850\um$ and the spectrum of the
far infra-red background to constrain our models, by ensuring that the
large quantities of cool dust invoked would not exceed the observed
emission.

We have shown that a variety of plausible obscured AGN models, which
provide good fits to the X-ray background spectrum and the number
counts at soft and hard X-ray energies, are consistent with the
observed sub-mm source counts and intensity of the FIRB.  The models
predict between 1 and 33 per cent of the FIRB intensity, and a similar
fraction of the number counts, depending on how extreme is the model,
but with the more conservative models predicting between 5 and 15 per
cent.  This is in good agreement with the fact that the majority of
sub-mm sources are identified as starburst galaxies, providing the
complementary sources.  In addition, the obscured AGN models may
be a suitable candidate for the source of the sub-mm emission
associated with EROs.  

Finally, we have made predictions of the redshift distribution of
sub-mm sources from our models, for both existing and proposed
surveys, and described how combined X-ray and sub-mm survey data will
be able to determine the extent of the obscured QSO contribution to
cosmological backgrounds at both high and low energies.

\section*{Acknowledgments}
 
This paper was prepared using the facilities of the STARLINK node at
Durham.  KFG acknowledges receipt of a PPARC studentship.  We thank
Chris Done and Ian Smail for useful discussions.  

\bibliographystyle{mnras}

\begin{thebibliography}{}

\bibitem[\protect\citeauthoryear{{Almaini}, {Lawrence}, \& {Boyle}}{{Almaini}
  et~al.}{1999}]{Almaini99}
{Almaini} O., {Lawrence} A.,  {Boyle} B.~J., 1999, \mnras, 305, L59

\bibitem[\protect\citeauthoryear{{Antonucci}}{{Antonucci}}{1993}]{Antonucci93}
{Antonucci} R., 1993, \araa, 31, 473

\bibitem[\protect\citeauthoryear{{Arnaud}}{{Arnaud}}{1996}]{Arnaud96}
{Arnaud} K.~A., 1996, in Jocoby G.H., Barnes J., eds, {\em Proc. of the ADASS V
  Conference}: ASP Conf. Series 101: San Francisco, Vol.~5, p.~17

\bibitem[\protect\citeauthoryear{{Barger} et~al.}{{Barger}
  et~al.}{1998}]{Barger98}
{Barger} A.~J., {Cowie} L.~L., {Sanders} D.~B., {Fulton} E., {Taniguchi} Y.,
  {Sato} Y., {Kawara} K.,  {Okuda} H., 1998, \nat, 394, 248

\bibitem[\protect\citeauthoryear{{Barger} et~al.}{{Barger}
  et~al.}{1999}]{Barger99}
{Barger} A.~J., {Cowie} L.~L., {Smail} I., {Ivison} R.~J., {Blain} A.~W.,
  {Kneib} J.~P., 1999, \aj, 117, 2656

\bibitem[\protect\citeauthoryear{{Beckett} et~al.}{{Beckett}
  et~al.}{1997}]{CIRSI}
{Beckett} M.~G., {Mackay} C.~D., {MCMahon} R.~G., {Parry} I.~R., {Piche} F.,
  {Ellis} R.~S., 1997, \procspie, 2871, 1152

\bibitem[\protect\citeauthoryear{{Benford} et~al.}{{Benford}
  et~al.}{1998}]{Benford98}
{Benford} D.~J., {Cox} P., {Omont} A.,  {Phillips} T.~G., 1998, AAS Meeting,
  192, 1104

\bibitem[\protect\citeauthoryear{{Benford} et~al.}{{Benford}
  et~al.}{1999}]{Benford99}
{Benford} D.~J., {Cox} P., {Omont} A., {Phillips} T.~G.,  {McMahon} R.~G.,
  1999, \apjl, 518, L65

\bibitem[\protect\citeauthoryear{{Blain}, {Ivison}, \& {Smail}}{{Blain}
  et~al.}{1998}]{Blain98}
{Blain} A.~W., {Ivison} R.~J.,  {Smail} I., 1998, \mnras, 296, L29

\bibitem[\protect\citeauthoryear{{Blain} et~al.}{{Blain}
  et~al.}{1999}]{Blain99}
{Blain} A.~W., {Kneib} J.~P., {Ivison} R.~J.,  {Smail} I., 1999, \apjl, 512,
  L87

\bibitem[\protect\citeauthoryear{{Boyle} et~al.}{{Boyle} et~al.}{1994}]{B94}
{Boyle} B.~J., {Shanks} T., {Georgantopoulos} I., {Stewart} G.~C.,  {Griffiths}
  R.~E., 1994, \mnras, 271, 639

\bibitem[\protect\citeauthoryear{{Campos} \& {Shanks}}{{Campos} \&
  {Shanks}}{1997}]{Campos97}
{Campos} A.,  {Shanks} T., 1997, \mnras, 291, 383

\bibitem[\protect\citeauthoryear{{Carilli} \& {Yun}}{{Carilli} \&
  {Yun}}{1999}]{Carilli99}
{Carilli} C.~L.,  {Yun} M.~S., 1999, \apjl, 513, L13

\bibitem[\protect\citeauthoryear{{Chapman} et~al.}{{Chapman}
  et~al.}{1999}]{Chapman99}
{Chapman} S.~C., {Scott} D., {Lewis} G.~F., {Borys} C.,  {Fahlman} G.~G., 1999,
  \mnras ~submitted. (astro-ph/9810444)

\bibitem[\protect\citeauthoryear{{Cimatti} et~al.}{{Cimatti}
  et~al.}{1997}]{Cimatti97}
{Cimatti} A., {Bianchi} S., {Ferrara} A.,  {Giovanardi} C., 1997, \mnras, 290,
  L43

\bibitem[\protect\citeauthoryear{{Comastri} et~al.}{{Comastri}
  et~al.}{1995}]{C95}
{Comastri} A., {Setti} G., {Zamorani} G.,  {Hasinger} G., 1995, \aap, 296, 1

\bibitem[\protect\citeauthoryear{{Dey} et~al.}{{Dey} et~al.}{1999}]{Dey99}
{Dey} A., {Graham} J.~R., {Ivison} R.~J., {Smail} I., {Wright} G.~S.,  {Liu}
  M., 1999, \apj, 519, 610

\bibitem[\protect\citeauthoryear{{Eales} et~al.}{{Eales}
  et~al.}{1999}]{Eales99}
{Eales} S., {Lilly} S., {Gear} W., {Dunne} L., {Bond} J.~R., {Hammer} F., {Le
  Fevre} O.,  {Crampton} D., 1999, \apj, 515, 518

\bibitem[\protect\citeauthoryear{{Efstathiou} \& {Rees}}{{Efstathiou} \&
  {Rees}}{1988}]{ER88}
{Efstathiou} G.,  {Rees} M.~J., 1988, \mnras, 230, 5P

\bibitem[\protect\citeauthoryear{{Fan}, {Strauss}, \& {SDSS
  Collaboration}}{{Fan} et~al.}{1998}]{hizqso}
{Fan} X., {Strauss} M.,  {SDSS Collaboration} , 1998, {\em SDSS Press Release}

\bibitem[\protect\citeauthoryear{{Fixsen} et~al.}{{Fixsen}
  et~al.}{1998}]{Fixsen98}
{Fixsen} D.~J., {Dwek} E., {Mather} J.~C., {Bennett} C.~L.,  {Shafer} R.~A.,
  1998, \apj, 508, 123

\bibitem[\protect\citeauthoryear{{Francis}}{{Francis}}{1993}]{Francis93}
{Francis} P.~J., 1993, \apj, 407, 519

\bibitem[\protect\citeauthoryear{{Frayer} et~al.}{{Frayer}
  et~al.}{1999}]{Frayer99}
{Frayer} D.~T. et~al., 1999, \apjl, 514, L13

\bibitem[\protect\citeauthoryear{{Frayer} et~al.}{{Frayer}
  et~al.}{1998}]{Frayer98}
{Frayer} D.~T., {Ivison} R.~J., {Scoville} N.~Z., {Yun} M.~S., {Evans} A.~S.,
  {Smail} I., {Blain} A.~W.,  {Kneib} J.~P., 1998, \apjl, 506, L7

\bibitem[\protect\citeauthoryear{{Gear} \& {Cunningham}}{{Gear} \&
  {Cunningham}}{1994}]{Gear94}
{Gear} W.~K.,  {Cunningham} C.~R., 1994, \procspie, 2198, 613

\bibitem[\protect\citeauthoryear{{Genzel}}{{Genzel}}{1997}]{FIRSTb}
{Genzel} R., 1997, in ASP Conf. Series 124: {\em Diffuse Infrared Radiation and
  the IRTS}, San Fransisco, p. 465

\bibitem[\protect\citeauthoryear{{Genzel} et~al.}{{Genzel}
  et~al.}{1998}]{Genzel98}
{Genzel} R. et~al., 1998, \apj, 498, 579

\bibitem[\protect\citeauthoryear{{Georgantopoulos} et~al.}{{Georgantopoulos}
  et~al.}{1997}]{ASCA1}
{Georgantopoulos} I., {Stewart} G.~C., {Blair} A.~J., {Shanks} T., {Griffiths}
  R.~E., {Boyle} B.~J., {Almaini} O.,  {Roche} N., 1997, \mnras, 291, 203

\bibitem[\protect\citeauthoryear{{Granato}, {Danese}, \&
  {Franceschini}}{{Granato} et~al.}{1997}]{Granato97}
{Granato} G.~L., {Danese} L.,  {Franceschini} A., 1997, \apj, 486, 147

\bibitem[\protect\citeauthoryear{{Gronwall} \& {Koo}}{{Gronwall} \&
  {Koo}}{1995}]{Gronwall95}
{Gronwall} C.,  {Koo} D.~C., 1995, \apjl, 440, L1

\bibitem[\protect\citeauthoryear{{Gunn}}{{Gunn}}{1999}]{KFG99PhD}
{Gunn} K.~F., 1999, PhD Thesis, University of Durham.

\bibitem[\protect\citeauthoryear{{Gunn} \& {Shanks}}{{Gunn} \&
  {Shanks}}{1999}]{KFG99}
{Gunn} K.~F.,  {Shanks} T., 1999, \mnras ~submitted.

\bibitem[\protect\citeauthoryear{{Haas} et~al.}{{Haas} et~al.}{1998}]{Haas98}
{Haas} M., {Chini} R., {Meisenheimer} K., {Stickel} M., {Lemke} D., {Klaas} U.,
   {Kreysa} E., 1998, \apjl, 503, L109

\bibitem[\protect\citeauthoryear{{Hauser} et~al.}{{Hauser}
  et~al.}{1998}]{Hauser98}
{Hauser} M.~G. et~al., 1998, \apj, 508, 25

\bibitem[\protect\citeauthoryear{{Hines}}{{Hines}}{1998}]{Hines98}
{Hines} D.~C., 1998, American Astronomical Society Meeting, 193, 2704

\bibitem[\protect\citeauthoryear{{Holland} et~al.}{{Holland}
  et~al.}{1999}]{Holland99}
{Holland} W.~S. et~al., 1999, \mnras, 303, 659

\bibitem[\protect\citeauthoryear{{Howarth}}{{Howarth}}{1983}]{Howarth83}
{Howarth} I.~D., 1983, \mnras, 203, 301

\bibitem[\protect\citeauthoryear{{Hu} \& {Ridgway}}{{Hu} \&
  {Ridgway}}{1994}]{Hu94}
{Hu} E.~M.,  {Ridgway} S.~E., 1994, \aj, 107, 1303

\bibitem[\protect\citeauthoryear{{Hughes} et~al.}{{Hughes}
  et~al.}{1998}]{Hughes98}
{Hughes} D.~H. et~al., 1998, \nat, 394, 241

\bibitem[\protect\citeauthoryear{{Ivison} et~al.}{{Ivison}
  et~al.}{1999}]{Ivison99}
{Ivison} R.~J., {Smail} I., {Blain} A.~W.,  {Kneib} J.~P., 1999, in
  preparation.

\bibitem[\protect\citeauthoryear{{Ivison} et~al.}{{Ivison}
  et~al.}{1998}]{Ivison98}
{Ivison} R.~J., {Smail} I., {Le Borgne} J.~F., {Blain} A.~W., {Kneib} J.~P.,
  {Bezecourt} J., {Kerr} T.~H.,  {Davies} J.~K., 1998, \mnras, 298, 583

\bibitem[\protect\citeauthoryear{{Lagache}}{{Lagache}}{1998}]{FIRBACK}
{Lagache} G., 1998, in {\em Wide Field Surveys in Cosmology}, 14th IAP meeting,
  Editions Frontieres, p. 301

\bibitem[\protect\citeauthoryear{{Laor} et~al.}{{Laor} et~al.}{1997}]{Laor97}
{Laor} A., {Fiore} F., {Elvis} M., {Wilkes} B.~J.,  {McDowell} J.~C., 1997,
  \apj, 477, 93

\bibitem[\protect\citeauthoryear{{Lawrence} et~al.}{{Lawrence}
  et~al.}{1993}]{Lawrence93a}
{Lawrence} A. et~al., 1993, \mnras, 260, 28

\bibitem[\protect\citeauthoryear{{Lewis} et~al.}{{Lewis}
  et~al.}{1998}]{Lewis98}
{Lewis} G.~F., {Chapman} S.~C., {Ibata} R.~A., {Irwin} M.~J.,  {Totten} E.~J.,
  1998, \apjl, 505, L1

\bibitem[\protect\citeauthoryear{{Madau} et~al.}{{Madau}
  et~al.}{1996}]{Madau96}
{Madau} P., {Ferguson} H.~C., {Dickinson} M.~E., {Giavalisco} M., {Steidel}
  C.~C.,  {Fruchter} A., 1996, \mnras, 283, 1388

\bibitem[\protect\citeauthoryear{{Madau}, {Ghisellini}, \& {Fabian}}{{Madau}
  et~al.}{1994}]{M94}
{Madau} P., {Ghisellini} G.,  {Fabian} A., 1994, \mnras, 270, L17

\bibitem[\protect\citeauthoryear{{Mather} et~al.}{{Mather}
  et~al.}{1994}]{Mather94}
{Mather} J.~C. et~al., 1994, \apj, 420, 439

\bibitem[\protect\citeauthoryear{{Metcalfe} et~al.}{{Metcalfe}
  et~al.}{1996}]{Metcalfe96}
{Metcalfe} N., {Shanks} T., {Campos} A., {Fong} R.,  {Gardner} J.~P., 1996,
  \nat, 383, 236

\bibitem[\protect\citeauthoryear{{Morrison} \& {McCammon}}{{Morrison} \&
  {McCammon}}{1983}]{MM83}
{Morrison} R.,  {McCammon} D., 1983, \apj, 270, 119

\bibitem[\protect\citeauthoryear{{Nandra} \& {Pounds}}{{Nandra} \&
  {Pounds}}{1994}]{NP94}
{Nandra} K.,  {Pounds} K.~A., 1994, \mnras, 268, 405

\bibitem[\protect\citeauthoryear{{Oliver} et~al.}{{Oliver}
  et~al.}{1998}]{Oliver98}
{Oliver} S. et~al., 1998, in {\em Wide Field Surveys in Cosmology}, 14th IAP
  meeting, Editions Frontieres, p. 165

\bibitem[\protect\citeauthoryear{{Pier} \& {Krolik}}{{Pier} \&
  {Krolik}}{1992}]{PK92b}
{Pier} E.~A.,  {Krolik} J.~H., 1992, \apj, 401, 99

\bibitem[\protect\citeauthoryear{{Pilbratt}}{{Pilbratt}}{1993}]{FIRSTa}
{Pilbratt} G., 1993, Advances in Space Research, 13, 912

\bibitem[\protect\citeauthoryear{{Puget} et~al.}{{Puget}
  et~al.}{1996}]{Puget96}
{Puget} J.~L., {Abergel} A., {Bernard} J.~P., {Boulanger} F., {Burton} W.~B.,
  {Desert} F.~X.,  {Hartmann} D., 1996, \aap, 308, L5

\bibitem[\protect\citeauthoryear{{Richards}}{{Richards}}{1999}]{Richards99}
{Richards} E.~A., 1999, \apjl, 513, L9

\bibitem[\protect\citeauthoryear{{Rigopoulou}, {Lawrence}, \&
  {Rowan-Robinson}}{{Rigopoulou} et~al.}{1996}]{Rigopoulou96}
{Rigopoulou} D., {Lawrence} A.,  {Rowan-Robinson} M., 1996, \mnras, 278, 1049

\bibitem[\protect\citeauthoryear{{Sanders} et~al.}{{Sanders}
  et~al.}{1989}]{Sanders89}
{Sanders} D.~B., {Phinney} E.~S., {Neugebauer} G., {Soifer} B.~T.,  {Matthews}
  K., 1989, \apj, 347, 29

\bibitem[\protect\citeauthoryear{{Schlegel}, {Finkbeiner}, \&
  {Davis}}{{Schlegel} et~al.}{1998}]{Schlegel98}
{Schlegel} D.~J., {Finkbeiner} D.~P.,  {Davis} M., 1998, \apj, 500, 525

\bibitem[\protect\citeauthoryear{{Seaton}}{{Seaton}}{1979}]{Seaton79}
{Seaton} M.~J., 1979, \mnras, 187, 73P

\bibitem[\protect\citeauthoryear{{Shaver} et~al.}{{Shaver}
  et~al.}{1996}]{Shaver96}
{Shaver} P.~A., {Wall} J.~V., {Kellermann} K.~I., {Jackson} C.~A.,  {Hawkins}
  M.~R.~S., 1996, \nat, 384, 439

\bibitem[\protect\citeauthoryear{{Smail}, {Ivison}, \& {Blain}}{{Smail}
  et~al.}{1997}]{SIB97}
{Smail} I., {Ivison} R.~J.,  {Blain} A.~W., 1997, \apjl, 490, L5

\bibitem[\protect\citeauthoryear{{Smail} et~al.}{{Smail} et~al.}{1998}]{SIBK98}
{Smail} I., {Ivison} R.~J., {Blain} A.~W.,  {Kneib} J.~P., 1998, in {\em After
  the dark ages: when galaxies were young (the Universe at $2<z<5$)}.
  (astro-ph/9810281)

\bibitem[\protect\citeauthoryear{{Smail} et~al.}{{Smail}
  et~al.}{1999}]{Smail99}
{Smail} I., {Ivison} R.~J., {Kneib} J.~P., {Cowie} L.~L., {Blain} A.~W.,
  {Barger} A.~J., {Owen} F.~N.,  {Morrison} G., 1999, \mnras ~in press.
  (astro-ph/9905246)

\bibitem[\protect\citeauthoryear{{Steidel} et~al.}{{Steidel}
  et~al.}{1999}]{Steidel99}
{Steidel} C.~C., {Adelberger} K.~L., {Giavalisco} M., {Dickinson} M.,
  {Pettini} M., 1999, \apj, 519, 1

\bibitem[\protect\citeauthoryear{{Tananbaum} et~al.}{{Tananbaum}
  et~al.}{1979}]{Tanan79}
{Tananbaum} H. et~al., 1979, \apjl, 234, L9

\bibitem[\protect\citeauthoryear{{Wang}}{{Wang}}{1991}]{Wang91}
{Wang} B., 1991, \apjl, 383, L37

\bibitem[\protect\citeauthoryear{{Werner}}{{Werner}}{1998}]{SIRTF}
{Werner} M.~W., 1998, American Astronomical Society Meeting, 193, 2502

\bibitem[\protect\citeauthoryear{{Yuan} et~al.}{{Yuan}
  et~al.}{1998}]{Yuan98_submmpaper}
{Yuan} W., {Brinkmann} W., {Siebert} J.,  {Voges} W., 1998, \aap, 330, 108

\bibitem[\protect\citeauthoryear{{Zheng} et~al.}{{Zheng}
  et~al.}{1997}]{Zheng97}
{Zheng} W., {Kriss} G.~A., {Telfer} R.~C., {Grimes} J.~P.,  {Davidsen} A.~F.,
  1997, \apj, 475, 469

\end{thebibliography}

\label{lastpage}

\end{document}